**Title**
Reproducible image-based profiling with Pycytominer

**Authors**
Erik Serrano[1], Srinivas Niranj Chandrasekaran[2], Dave Bunten[1], Kenneth I. Brewer[3], Jenna Tomkinson[1], Roshan Kern[1,4], Michael Bornholdt[2], Stephen Fleming[5], Ruifan Pei[2], John Arevalo[2], Hillary Tsang[2], Vincent Rubinetti[1], Callum Tromans-Coia[2], Tim Becker[2], Erin Weisbart[2], Charlotte Bunne[2], Alexandr A. Kalinin[2], Rebecca Senft[2], Stephen J. Taylor[1], Nasim Jamali[2], Adeniyi Adeboye[2], Hamdah Shafqat Abbasi[2], Allen Goodman[2,6], Juan C. Caicedo[2,7], Anne E. Carpenter[2], Beth A. Cimini[2], Shantanu Singh[2,*], Gregory P. Way[1,*]

***** Corresponding Authors: shsingh@broadinstitute.org; gregory.way@cuanschutz.edu

**Affiliations**
1. Department of Biomedical Informatics, University of Colorado School of Medicine
2. Imaging Platform, Broad Institute of MIT and Harvard
3. Independent Researcher
4. Case Western Reserve University
5. Data Sciences Platform, Broad Institute of MIT and Harvard
6. Genentech gRED
7. Morgridge Institute for Research, University of Wisconsin-Madison

**Abstract**
Advances in high-throughput microscopy have enabled the rapid acquisition of large numbers of high-content microscopy images. Whether by deep learning or classical algorithms, image analysis pipelines then produce single-cell features. To process these single-cells for downstream applications, we present Pycytominer, a user-friendly, open-source python package that implements the bioinformatics steps, known as “image-based profiling”. We demonstrate Pycytominer’s usefulness in a machine learning project to predict nuisance compounds that cause undesirable cell injuries.

**Main**
In the past thirty years, high-content microscopy has undergone a remarkable technological transformation that has given scientists the ability to acquire thousands of single-cell measurements in high-throughput experiments.[1] In turn, open-source image analysis software has proliferated, including Fiji[2], CellProfiler[3], and others. These tools can derive biological insights from large microscopy datasets, but they lack functions for downstream bioinformatics processing of image-based features. Recently, the field of image-based profiling has emerged which requires this kind of processing to accomplish various useful biological applications.[4,5] Thus far, the primary applications for image-based profiling have been in drug development.[4,6] Specifically, image-based profiling offers disease phenotype discovery, target identification, drug repurposing, toxicity assessment, and the exploration of novel therapeutic hypotheses.[7] Image-based profiling also enables innovative studies into fundamental biological processes such as cell death[8,9], grouping functional genes[10,11], and mitochondrial dynamics.[12]

To generate image-based profiles, scientists first prepare cell samples that can be subjected to small molecule or genetic perturbations, where, after an incubation period, they undergo fluorescence staining to mark specific cellular compartments, followed by microscopy imaging **(Figure 1A)**.[4] During image analysis, scientists perform quality control, cell segmentation, and high-content single-cell feature extraction. Feature extraction algorithms (including emerging approaches based on deep learning) quantify morphological properties per cell, such as shapes, sizes, stain intensities and more (**Figure 1A**). Image-based profiling then processes these features to prepare them for downstream analyses (**Figure 1B**).

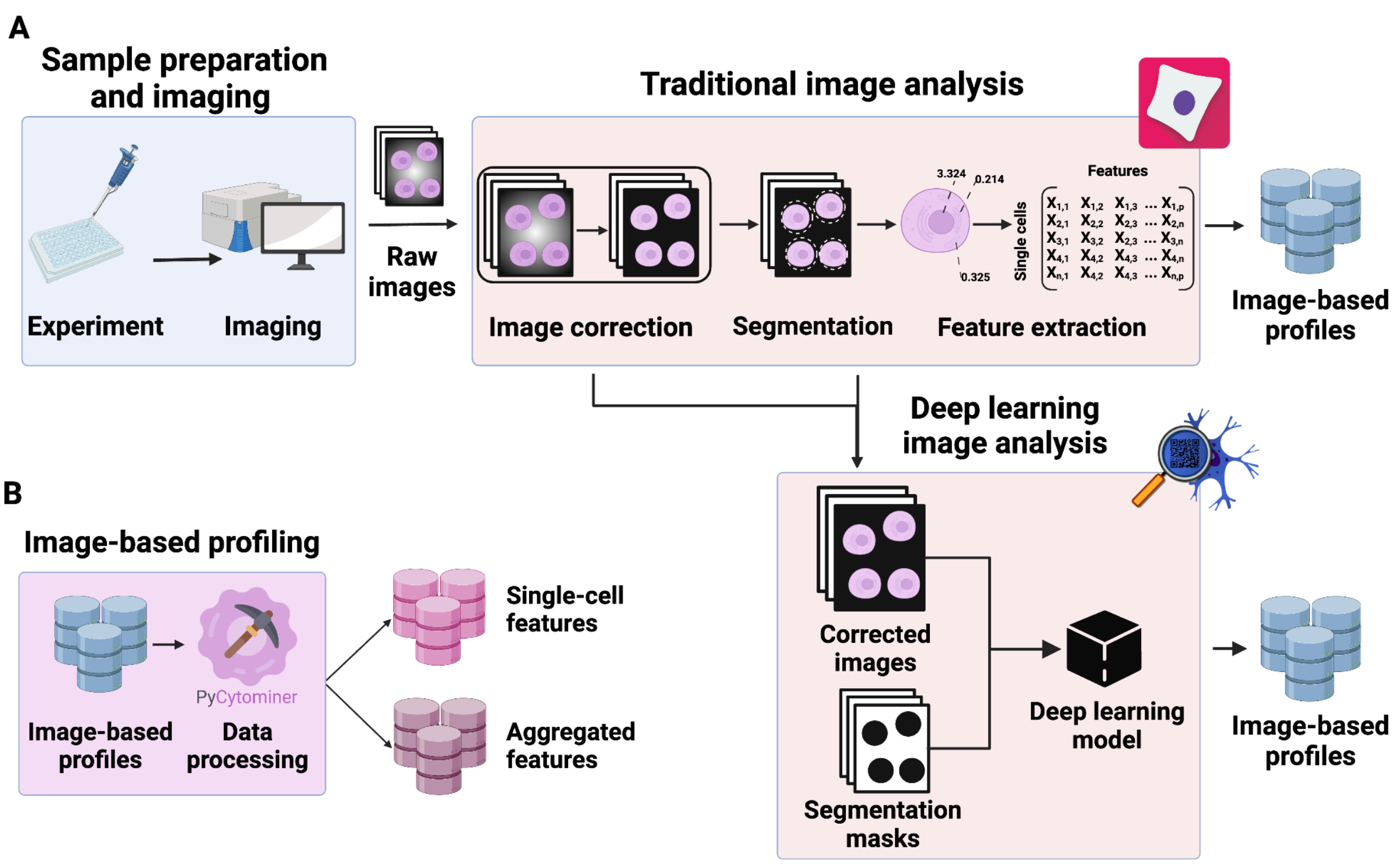

**Figure 1. The standard image-based profiling experiment and the role of Pycytominer. (A)** In the experimental phase, a scientist plates cells, often perturbing them with chemical or genetic agents and performs microscopy imaging. In image analysis, using CellProfiler for example, a scientist applies several data processing steps to generate image-based profiles. In addition, scientists can apply a more flexible approach by using deep learning models, such as DeepProfiler, to generate image-based profiles. **(B)** Pycytominer performs image-based profiling to process morphology features and make them ready for downstream analyses.

Here we introduce Pycytominer, an open-source Python[13] package for designing and running image-based profiling bioinformatics pipelines. This project resulted from discussions and consensus stemming from the community's inaugural work together.[5] We developed Pycytominer using Pandas[14], Apache Parquet[15], and SQLAlchemy[16], and it offers a modular Application Programming Interface (API) to enable flexible customization. The Pycytominer API performs the core functions—aggregate, annotate, normalize, and feature selection—and offers several helpful utility functions including batch effect correction (**Supplementary Figure 1A**).

Pycytominer distinguishes itself from related image-based profiling software tools (see **Supplementary Table 1** for a comparison) by providing a comprehensive, user-friendly, and actively maintained solution in the widely-used Python programming language, with a modular design and seamless integration with popular data processing libraries. Pycytominer has already been used in many applications. Most notably, Pycytominer processed two of the largest publicly-available high-content microscopy datasets, Joint Undertaking in Morphological Profiling (JUMP)[17] and Library of Integrated Network-Based Cellular Signatures (LINCS).[10] Pycytominer has also processed the majority of the currently 31 Cell Painting datasets in the Cell Painting Gallery.[18] We encourage Pycytominer users to look at these existing resources for example pipelines, but we also provide full tutorials, user documentation, a handbook, an orchestration recipe, and comprehensive walkthroughs (**Supplementary Figure 2**; links in the **Online Methods**).

To showcase Pycytominer usage, we reprocessed a publicly-available high-content microscopy dataset from a study aimed at discerning nuisance compounds that induce undesirable cell injuries (**Figure 2A**).[19] Downstream of Pycytominer, we trained a multi-class logistic regression model to predict 15 cell injury categories. We split the data into training and testing sets, and included three different types of holdout data: Whole plate, whole treatments, and individual wells (**Supplementary Table 2**). Overall, our model demonstrated strong performance in identifying most cellular injury types including “cytoskeletal”, “HSP90”, and “HDAC” with high F1 scores **(Figure 2B)**. Notably, other cell injuries such as “tannin,” “saponin,” and “nonspecific reactive” showed relatively low performance. Overall, predictions were consistently strong, even in wells and plates held out from training **(Figure 2C)** and especially when compared to shuffled baseline **(Supplementary Figure 3)**. When holding out individual treatments, we observed high performance for predicting “genotoxin” and “redox” injuries, but poor performance for “saponin” and “kinase”, which further demonstrates differences in generalizability across injury types (**Supplementary Figure 4**).

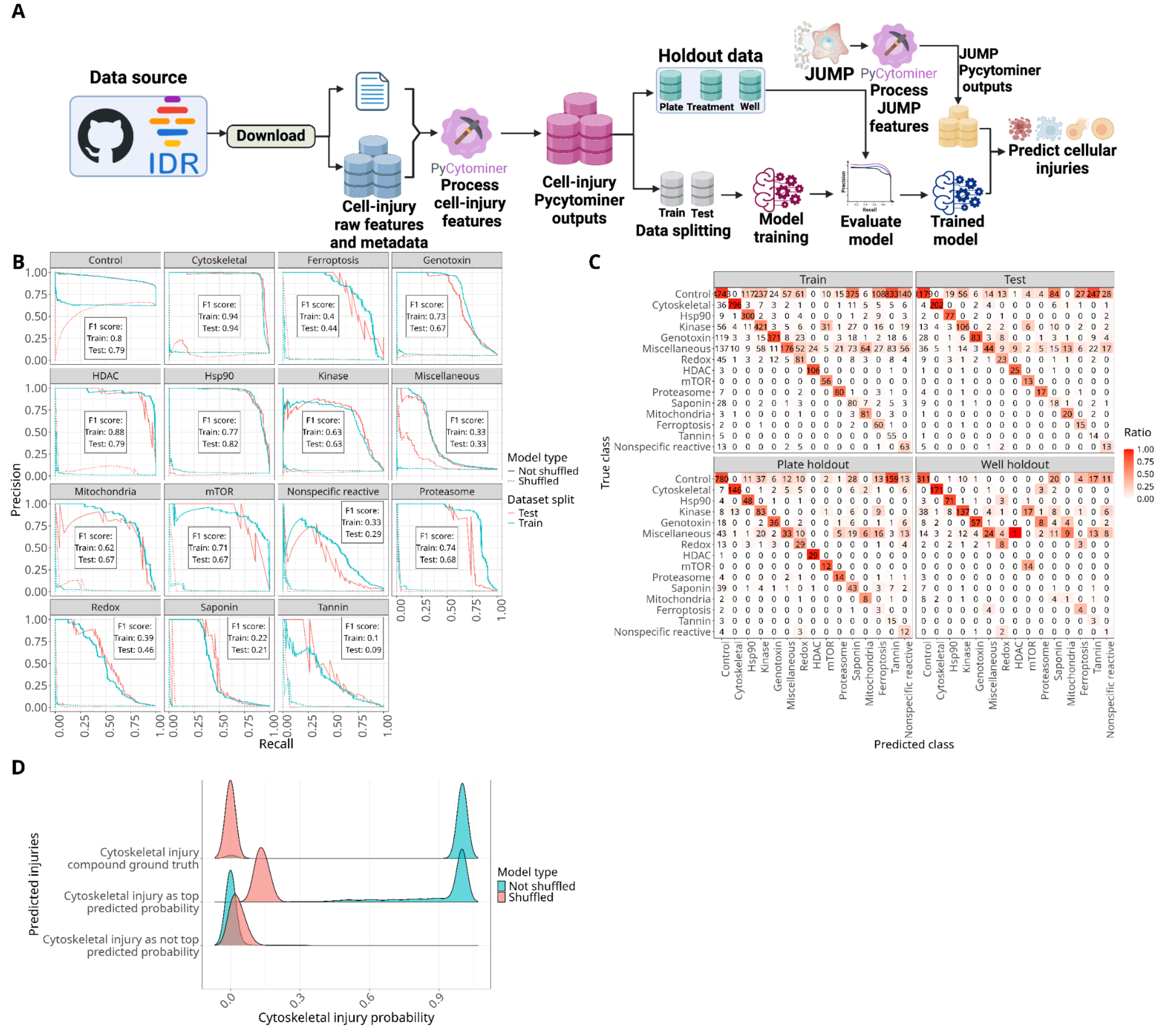

**Figure 2. Model performance and evaluation with JUMP data. (A)** Our pycytominer-based workflow to process publicly-available data and to train a machine learning model to predict cellular injury. **(B)** Precision and recall scores for predicting various cellular injuries, comparing the not shuffled model (solid lines) with the shuffled model (dashed lines) across distinct injury types and data splits, with blue indicating the training set and red indicating the test set. The F1 scores for each injury represent only the testing and training datasets with the not shuffled model. **(C)** Confusion matrices assessing the model's predictive performance across training, testing, and holdout data. **(D)** Cytoskeletal injury probability distribution generated by the shuffled (red) and non-shuffled (blue) models. The three groups represent the ground truth wells (top), wells predicted to have cytoskeletal injury (middle), and wells not predicted to have cytoskeletal injury (bottom).

We then applied our pre-trained cell injury model to the CPJUMP1 pilot dataset, which comprises 38,996 well-level profiles from cell populations treated with 636 genetic and chemical perturbations.[17] We also processed CPJUMP1 using Pycytominer, and we identified 24 wells overlapping with cytoskeletal injury compounds in the training dataset. Our model accurately predicted 23 out of these 24 JUMP ground truth wells **(Figure 2D)**. Importantly, the model predicted many other non-labeled compounds to have cytoskeletal injuries with high probability, while most other compounds showed lower probabilities. We observed similar prediction trends with the other cellular injury classes **(Supplementary Figure 5)**. We present injury type predictions for all CPJUMP1 chemical and genetic perturbations in **Supplementary Table 3**.

Nevertheless, Pycytominer has certain limitations. Being written in Python may exclude users proficient in other programming languages and necessitates integrating their analytical pipelines into Python. To address this, our future roadmap includes more containerization and the addition of command line interface (CLI) options, which will broaden access and offer multilingual support. Second, there may be more optimal image-based profiling methods not yet discovered. In anticipation of these future developments, we have designed the Pycytominer API and testing framework with modularity in mind. This approach allows for the easy incorporation of new methods as they surpass the current state of the art. Third, it is important to note that Pycytominer focuses on a specific segment of the entire image analysis pipeline, beginning with image analysis outputs and ending with fully-processed single-cell and bulk image-based profiles. Consequently, users are required to be proficient in other software for preliminary processing steps like quality control, segmentation, and feature extraction as well as downstream tools for analysis and visualization. This decision to concentrate on core image-based profiling functionality simplifies software maintenance and fosters direct innovation in this area.

Looking to the future, Pycytominer is poised to play an essential role as an integral tool for image-based profiling. With a steadfast commitment and a growing community consistently contributing new and optimized functionality, Pycytominer offers a reliable and standardized toolkit that empowers researchers to unveil new insights in multiple fields from drug discovery to fundamental cell biology research.

## Online Methods

### *Pycytominer software practices*

We rigorously apply open-source best practices during Pycytominer's development in four main categories: Implementation, testing, release, and community (**Supplementary Figure 1B**). (1) *Implementation*. Pycytominer eases the process of contributing code by providing development container specifications usable in VSCode or GitHub Codespaces that contain the full set of software dependencies needed to develop and test the codebase. When changes are ready, contributors submit pull requests, which must be reviewed to ensure adherence to best practices such as modularization, code styling, and documentation. (2) *Testing*. Pycytominer's comprehensive testing suite, including unit tests and code coverage analysis, serves as a crucial step to ensure the correctness and functionality of the software implementation. Testing every new change against the full test suite reduces the introduction of software bugs and ensures consistent behavior across versions. (3) *Release*. Pycytominer follows semantic versioning and maintains a changelog to ensure users are kept informed of new features and important changes. Releases are made available directly on GitHub and are also packaged for use within Python's two major package repositories PyPI and conda. In addition, Pycytominer supports operating environment containerization (facilitated by Docker[20]), encapsulating dependencies to enhance reproducibility. (4) *Community*. Pycytominer cultivates our open-source community by welcoming new contributors with clear contributing instructions and guidelines and a code-of-conduct to ensure professional standards are kept. These community efforts are essential for good collaboration, maintaining quality, and ensuring project sustainability. Embracing the full set of these best software practices fosters a collaborative environment that facilitates continuous improvements, encourages reproducibility, welcomes newcomers, and contributes to package usability for developers and users alike.

### Code/data availability and tutorials

- Pycytominer is an open-source project and its source code can be viewed and downloaded from: https://github.com/cytomining/pycytominer
- Pyctominer's installation and usage documentation is available at: https://Pycytominer.readthedocs.io/
- A tutorial on how to conduct single-cell image-based profiling is available at: https://Pycytominer.readthedocs.io/en/latest/walkthroughs/single_cell_usage.html
- Repository containing the code used to conduct analysis and generate results: https://github.com/WayScience/Cytotoxic-Nuisance-Metadata-Analysis

### Reprocessing the cell injury dataset

The cell injury profiles used in this study were generated by Dahlin et al.[19] and consisted of image-based profiles along with associated metadata. These profiles were derived from U-2OS cells treated with 218 cytotoxic compounds in a concentration-response format. The dataset, accession number 0133, was made publicly available on the Image Data Resource (IDR). The *cpg0000-jump-pilot* dataset, also referred to as CPJUMP1, was accessed via the Cell Painting Gallery website, specifically from the path 'cpg0000-jump-pilot/source_4/workspace/profiles/2020_11_04_CPJUMP1'. We downloaded only the normalized aggregated profiles that have been normalized to their negative controls, resulting in a total of 51 normalized plate profiles. We used Pycytominer's feature selection function to select informative morphological features to include during model training. Specifically, we utilized the variance threshold operation to filter out features with low variance across the dataset. Features with a low frequency of 0.05 and below were removed due to having large differences between most common features within the dataset. Additionally, we incorporated an additional parameter called 'na_cutoff,' which we set to 0.05, indicating that features would be dropped if the proportion of missing values (NaN) exceeded 5%. This resulted in selecting 221 morphological features for model training.

### Data labeling

The cell injury dataset contained 23,111 wells. A filtering process was conducted to select wells with treatments known to cause cellular injury. We split the dataset into control and treated groups, resulting in 9,855 control wells and 6,484 treated wells. Wells treated with DMSO were labeled as "Control," while the treated wells were labeled according to their specific treatments. We focused on wells treated with chemical perturbants identified in the cell injury study, which included 15 types of injuries: Control, Cytoskeletal, Hsp90, Kinase, Genotoxin, Miscellaneous, Redox, HDAC, mTOR, Proteasome, Saponin, Mitochondria, Ferroptosis, Tannin, and Nonspecific Reactive. Wells that did not contain a known treatment with associated injury were not considered, therefore resulted in a total of 16,703 wells for our study.

### Identifying shared features

We identified shared features by extracting feature names from both the cell injury and CPJUMP1 datasets. Treating the feature names as sets allowed us to generate a unified set identifying the intersecting features present in both datasets. These intersecting features represent the shared features between the two datasets. Utilizing this shared feature space, we conducted feature selection and model training.

**Identifying shared treatments**
We used International Chemical Identifiers (InChIKeys) to identify chemical perturbants in both datasets. In CPJUMP1, 24 wells contained chemical perturbants labeled as causing cytoskeletal injuries, as indicated in the cell injury study. We set these wells as our ground truth when conducting our evaluation.

**Machine learning model training**
We first generated three holdout datasets in the specific sequence: plate, treatment, and well holdouts. We created plate holdouts by randomly selecting 10 plates from the feature-selected dataset. For treatment holdouts, we identified specific cell injuries with more than 10 associated treatments; we excluded injuries that did not meet this criterion from the holdout. We generated well holdouts by selecting all non-heldout plates and randomly choosing 15 wells (5 controls and 10 wells with each cell injury) from each plate. The remaining data (13,502 wells) constituted our training dataset, which we further split into 80% (10,801 wells) for training and 20% (2,701 wells) for testing, maintaining the same class label proportions to address dataset imbalance.

We trained and hyperparameterized our multi-class logistic regression model using scikit-learn's RandomizedSearchCV. To address label imbalance, we configured our logistic regression model to automatically adjust class weights based on their frequencies. This adjustment helps to mitigate the impact of imbalanced labels during model training. For the hyperparameter tuning, we defined a parameter grid for RandomizedSearchCV. This grid included exploration of three different penalties: Lasso (L1), Ridge (L2), and Elastic Net (combination of L1 and L2). We varied the regularization strength across a range of values (0.0001, 0.01, 0.1, 1, 10, 100) to assess its impact on model performance. Additionally, we experimented with different tolerance values (1e-6 and 1e-3) to determine the threshold for stopping the hyperparameter search process.

To explore the Elastic Net regularization further, we searched different ratios (0.1, 0.3, 0.5, 0.7, 0.9) between L1 and L2 penalties. Finally, we evaluated various solvers (newton-cg, lbfgs, liblinear, sag, saga) to optimize the logistic regression model parameters through RandomizedSearchCV. Each solver offers distinct optimization techniques, allowing us to assess their effectiveness in our model's context.

**Acknowledgements**
We would like to thank Michael Lippincott for performing code review.

This software was created with the support of NIH grant R35 GM122547 to AEC and institutional startup funding to GPW. BAC was supported by grant 2020-225720 (DOI:10.37921/977328pjvbca) from the Chan Zuckerberg Initiative DAF, an advised fund of Silicon Valley Community Foundation (funder DOI 10.13039/100014989). Research reported in this publication was supported by the National Library of Medicine of the National Institutes of Health (NIH) under award number T15LM009451 to ES. The authors gratefully acknowledge funding from the Massachusetts Life Sciences Center Bits to Bytes Capital Call program (to AEC) and the Data Science Internship Program (supporting RP and AA), partners of the JUMP Cell Painting Consortium, Starr Cancer Consortium (112-0039 to AEC), Schmidt Fellowship program of the Broad Institute (to JCC), and the Human Frontier Science Program (RGY0081/2019 to SS).

**Disclosure of competing interests**
The authors declare they have no competing interests relevant to this work.

**Lead contacts**
For further information and resource requests, please direct your inquiries to Gregory Way (gregory.way@cuanschutz.edu) and Shantanu Singh (shantanu@broadinstitute.org).

## Supplementary information

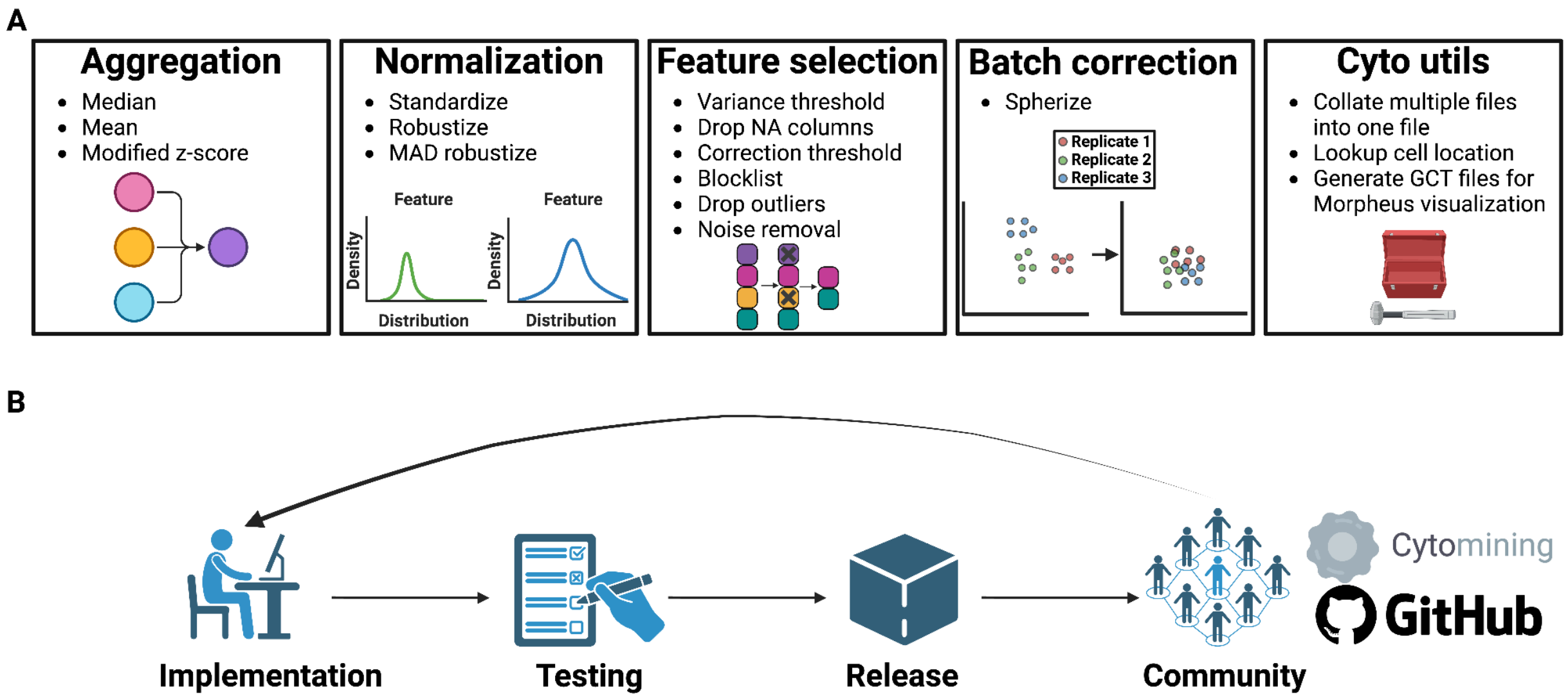

**Supplementary Figure 1.** *Pycytominer's core Application Programming Interface (API) and software practices.*

(**A)** Pycytominer performs five fundamental functions, each implemented with a simple and intuitive API. Each function enables a user to implement various methods for executing operations. **(B)** The open-source Pycytominer community supports best development practices to facilitate code effectiveness and longevity.

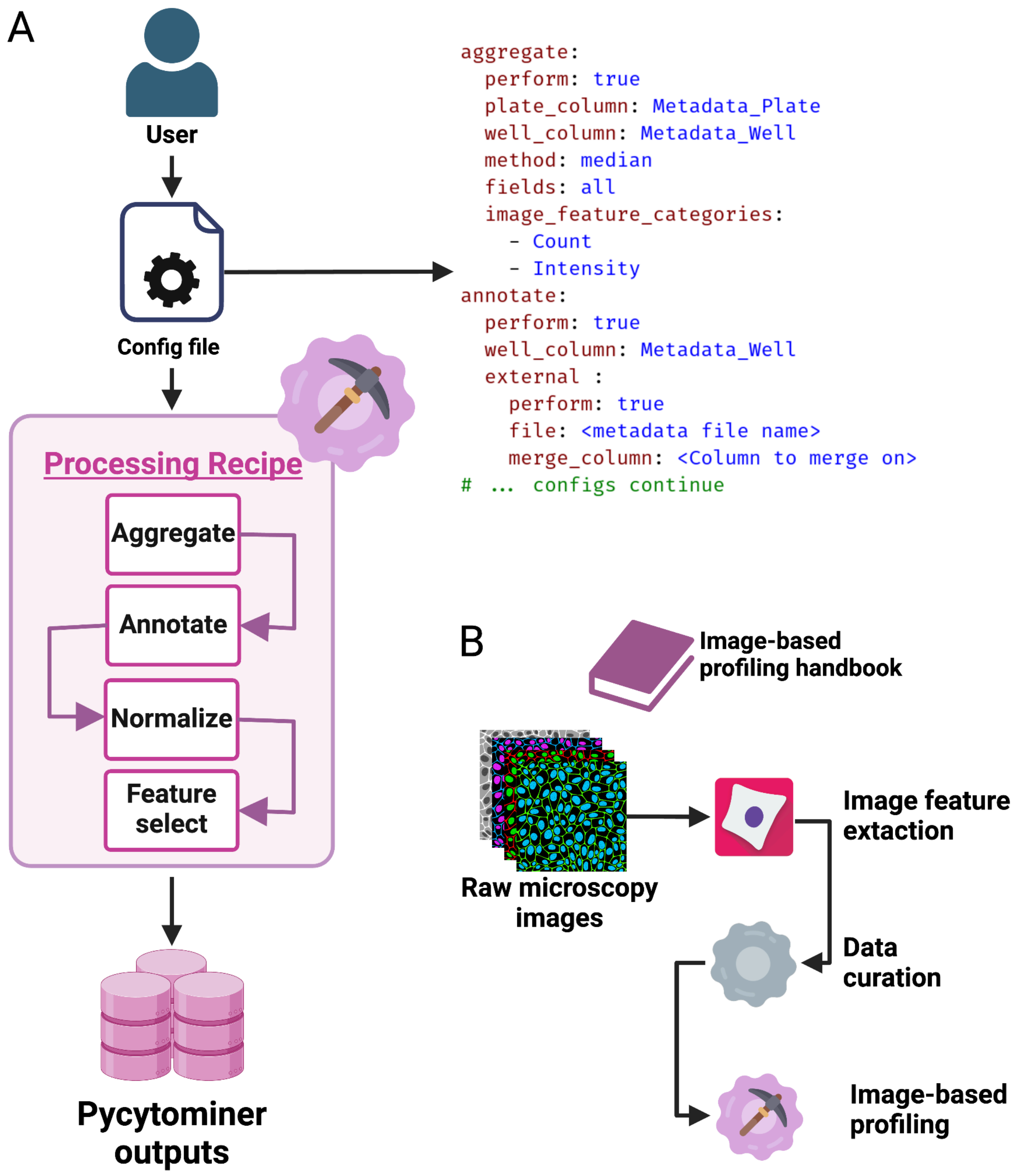

**Supplementary Figure 2.** *Pycytominer recipe and handbook for image-based profiling.*

**(A)** Users can configure a profiling recipe to customize Pycytominer implementation of image-based profiling steps. Users interact with the profiling recipe through a configuration file provided in yaml format, to parameterize each function within the Pycytominer workflow. **(B)** We have also written an image-based profiling handbook available at https://cytomining.github.io/profiling-handbook/, which documents all steps in a full image analysis workflow, which

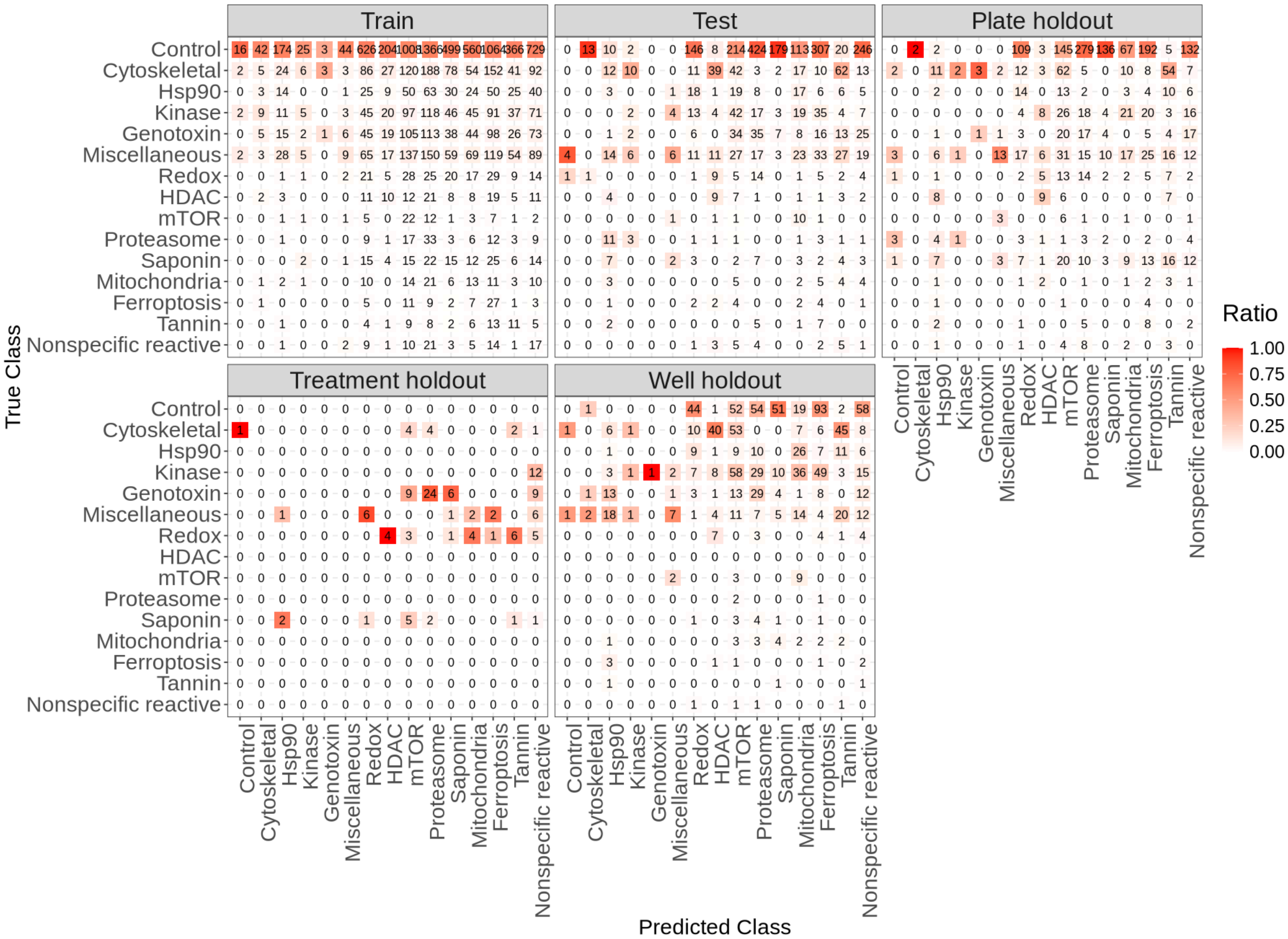

**Supplementary Figure 3.** *Shuffled multi-class logistic regression model predictive performance.*

The confusion matrix demonstrates the model's ability to predict cell injuries across various data splits and holdouts. The red color gradient represents the ratio of the count over the true class label.

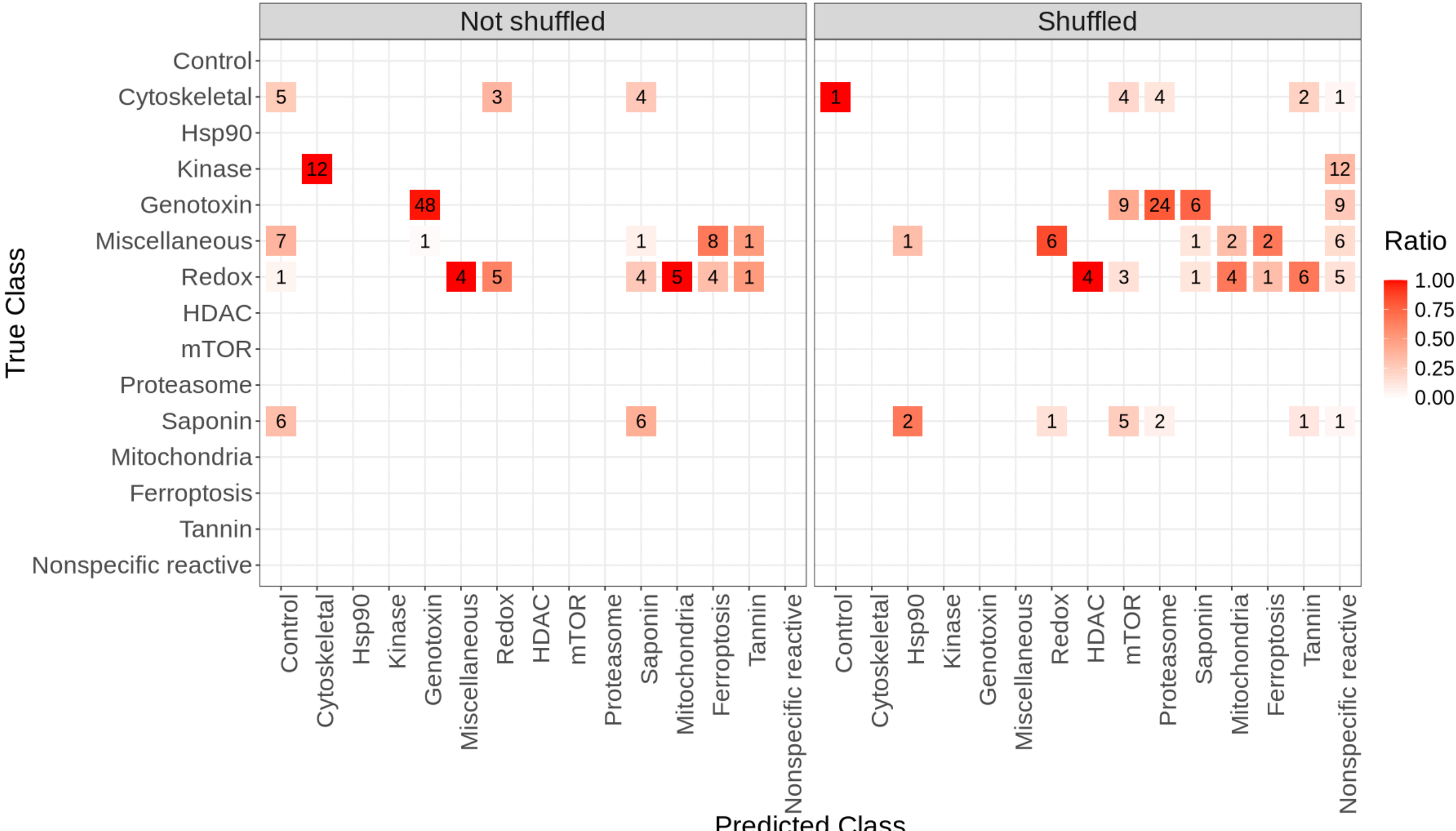

**Supplementary Figure 4.** *Comparison of multi-class logistic regression model predictive performance for treatment holdouts.*

The confusion matrix evaluates the predictive performance of the Not Shuffled model (left) versus the Shuffled model (right) using only the treatment holdout dataset. The red color gradient corresponds to the ratio, indicating the count over the true class label.

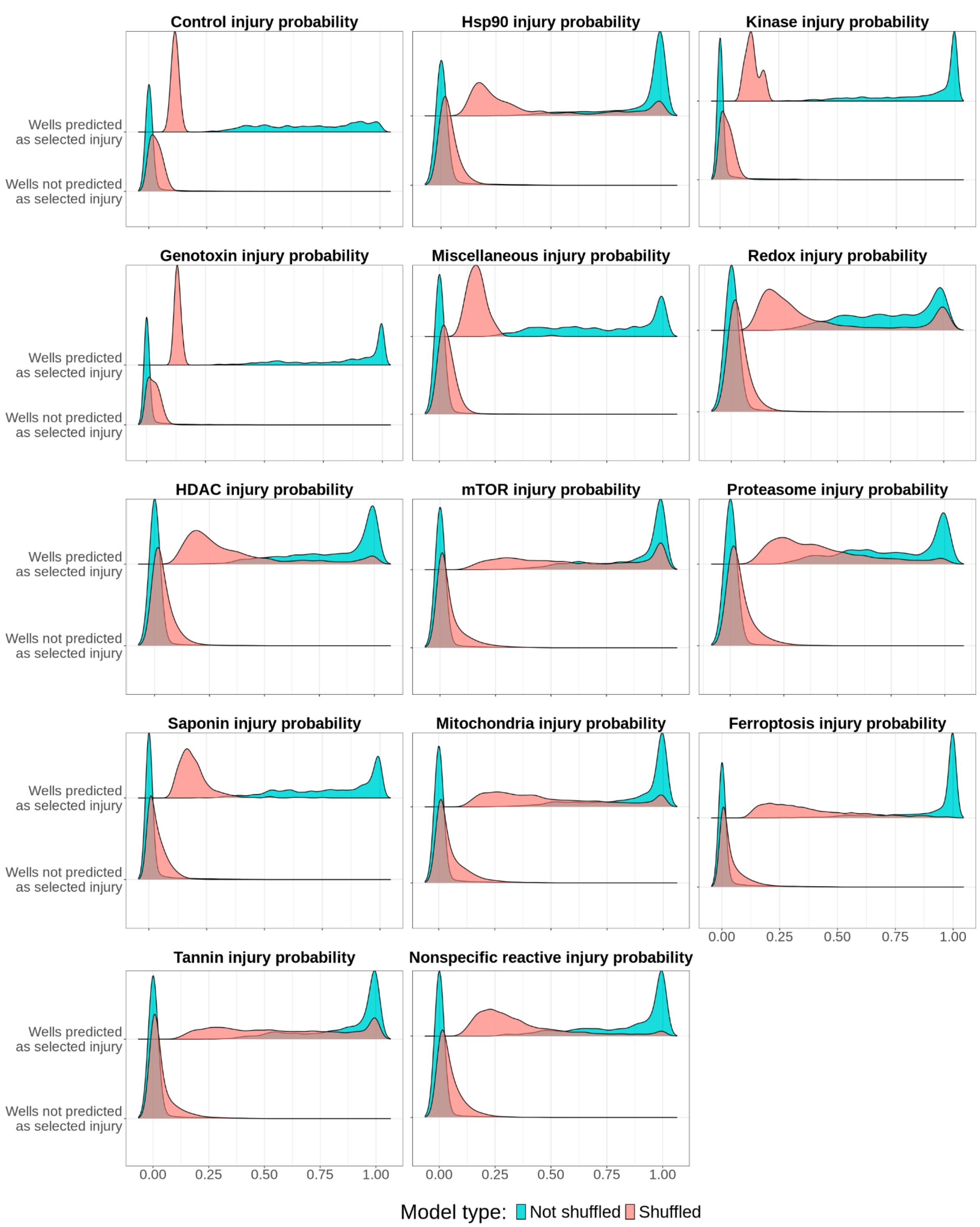

**Supplementary Figure 5.** *Predicting all other cellular injury types in JUMP data.*

Probability distributions generated by our model for predicting each injury type in the Joint Undertaking in Morphological Profiling (JUMP) Pilot data (CPJUMP1). Per facet, the upper curves indicate wells predicted to have the selected injury (with the top probability), while the lower curves indicate wells not predicted to have the selected injury (anything but top probability). The shuffled model shows lower top probabilities and higher other probabilities compared to the not shuffled model, which shows high model confidence for most injury types, similar to our ground truth performance (see **Figure 2D**).

| Software | Language | Key Features | Limitations |
| --- | --- | --- | --- |
| Pycytominer | Python | Comprehensive functionality for image-based profiling.<br><br>User-friendly API and documentation.<br><br>Actively maintained and supported.<br><br>Integrates with popular data processing libraries. | Requires proficiency in Python for integration with workflows.<br><br>Pycytominer focuses on image-based profiling, thus necessitating both upstream image analysis and downstream analyses. |
| BioProfiling.jl | Julia | Implements best practices for image-based profiling.<br><br>Fast performance due to Julia's compilation. | Limited user base and community support compared to Python. |
| StratoMineR | Web-based | Designed for analyzing high-content screening experiments.<br><br>Graphical user interface for ease of use. | Commercial tool with limited customization options. |
| Squidpy | Python | Specialized for spatial data analysis.<br><br>Integrates with scanpy for single-cell analysis. | Focuses on cell-type identification and gene expression.<br><br>Limited functionality for general image-based profiling. |
| PhenoRipper | MATLAB | Rapid feature extraction and data visualization.<br><br>Designed for high-content screening data. | Requires MATLAB; last official update was in 2011.<br><br>Very little options to process image-based profiles. |

**Supplementary Table 1.** *Overview of image-based profiling software tools*

Well-known image-based profiling software tools, highlighting their key features and limitations.

| Cellular Injury | Total Wells | Training Wells | Testing Wells | Plate Holdout Wells | Treatment Holdout Wells | Well Holdout Wells |
|---|---|---|---|---|---|---|
| Control | 9855 | 6726 | 1682 | 1072 | 0 | 375 |
| Cytoskeletal | 1472 | 881 | 221 | 181 | 12 | 177 |
| Ferroptosis | 96 | 66 | 16 | 6 | 0 | 8 |
| Genotoxin | 944 | 590 | 147 | 73 | 48 | 86 |
| HDAC | 168 | 110 | 28 | 30 | 0 | 0 |
| Hsp90 | 552 | 334 | 84 | 54 | 0 | 80 |
| Kinase | 1104 | 600 | 150 | 120 | 12 | 222 |
| Miscellaneous | 1304 | 806 | 201 | 172 | 18 | 107 |
| Mitochondria | 144 | 92 | 23 | 12 | 0 | 17 |
| Nonspecific reactive | 128 | 84 | 21 | 19 | 0 | 4 |
| Proteasome | 144 | 94 | 23 | 24 | 0 | 3 |
| Redox | 312 | 172 | 43 | 54 | 24 | 19 |
| Saponin | 288 | 131 | 33 | 102 | 12 | 10 |
| Tannin | 96 | 60 | 15 | 18 | 0 | 3 |
| mTOR | 96 | 56 | 14 | 12 | 0 | 14 |

**Supplementary Table 2.** *Overview of data splitting process for cell injury dataset*

We categorized the cell injury dataset into distinct subsets for model training, all stemming from the "Total Wells" category (green). "Train Wells" and "Test Wells" were used for model training and testing, respectively (orange). Additionally, we used three holdout sets (pink) to evaluate model performance in never-before-seen plates, treatments, and wells.

**Supplementary Table 3.** *Cell injury model prediction and probability scores for the JUMP-CP dataset*

This table contains all the predictions made by our model for each well in the CPJUMP1 dataset. Each prediction is accompanied by a probability score indicating its confidence level.